\pdfoutput=1 
\documentclass{JINST}

\usepackage{amsmath, amssymb, bm}
\usepackage{fancyhdr}
\usepackage{placeins}
\usepackage{graphicx}
\usepackage{wrapfig}
\usepackage[small]{caption}
\usepackage[labelformat=simple]{subcaption}

\title{Photon Detection System Designs for the Deep Underground Neutrino Experiment}

\author{ D.~Whittington\\
         Physics Department, Indiana University,\\
         Bloomington, IN, USA\\
         E-mail: \email{dwwhitti@indiana.edu} }
\date{November 11, 2015}
\preprint{FERMILAB-CONF-15-490-ND-PPD}

\abstract{The Deep Underground Neutrino Experiment (DUNE) will be a premier facility for exploring long-standing questions about the boundaries of the standard model. Acting in concert with the liquid argon time projection chambers underpinning the far detector design, the DUNE photon detection system will capture ultraviolet scintillation light in order to provide valuable timing information for event reconstruction. To maximize the active area while maintaining a small photocathode coverage, the experiment will utilize a design based on plastic light guides coated with a wavelength-shifting compound, along with silicon photomultipliers, to collect and record scintillation light from liquid argon. This report presents recent preliminary performance measurements of this baseline design and several alternative designs which promise significant improvements in sensitivity to low-energy interactions.}


\begin{document}

\section{Introduction}\label{sec:intro}

The Deep Underground Neutrino Experiment (DUNE) is a proposed long-baseline accelerator-based neutrino oscillation experiment~\cite{bib:DUNE-Overview}. A neutrino beam originating at Fermi National Accelerator Laboratory (Fermilab) as part of the Long-Baseline Neutrino Facility will be directed toward a far detector site at the Sanford Underground Research Facility in Lead, South Dakota. The DUNE far detector will consist of four 10-kT active volume liquid argon time projection chambers (TPC). The goal of the experiment is to measure electron neutrino appearance and muon neutrino disappearance probabilities over a range of energies and a baseline of 1,300 km.

The photon detection system of the DUNE single-phase liquid argon TPC far detector will employ a light guide-based design. This paradigm will accommodate a large UV-sensitive active surface area with reduced photocathode area and a slim profile to fit inside the wrapped anode plane assembly design~\cite{bib:DUNE-FarDet}. Liquid argon (LAr) is a copious source of scintillation light in the vacuum ultraviolet (VUV, specifically 128~nm), so a VUV-sensitive detector is required. This scintillation signal will be a valuable timing measurement to inform TPC event reconstruction and provide a trigger for non-beam events such as proton decay or supernova burst neutrino interactions.

\begin{figure}[ht]
  \begin{center}
    \includegraphics[width=0.7\columnwidth]{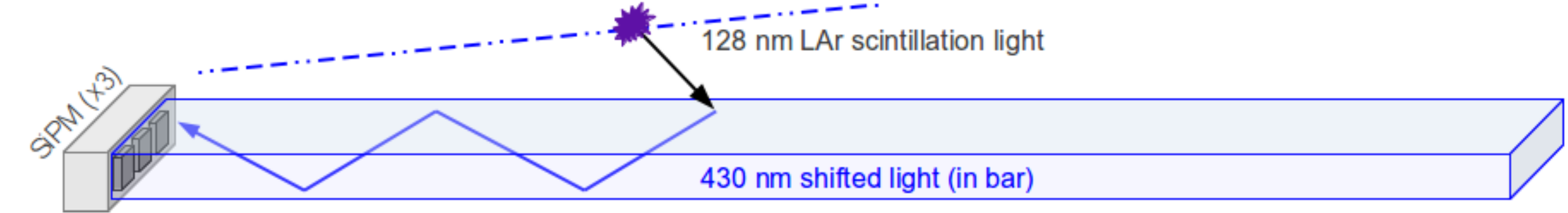}
    \caption{Baseline light guide detector design.}
    \label{fig:BaselineLightGuide}
  \end{center}
  \vspace{-1em}
\end{figure}

Figure~\ref{fig:BaselineLightGuide} illustrates the baseline light guide design for DUNE. It consists of an acrylic bar with a surface layer of wavelength shifter (WLS) such as TPB or bis-MSB to convert this 128~nm signal to visible blue photons and transport them via total internal reflection to a readout device~\cite{bib:MIT-LightGuide}. Silicon photomultipliers (SiPM) are low-noise, slim-profile, and low-voltage devices which are a promising choice for the detection of the optical signal. In order to improve on the attenuation length of the baseline design several alternative designs were tested in this experiment which decouple the VUV conversion from transport of the optical signal.

\FloatBarrier
\section{Facility and Apparatus}

\begin{wrapfigure}{r}{0.35\columnwidth}
  \vspace{-2em}
  \begin{center}
    \includegraphics[width=0.35\columnwidth]{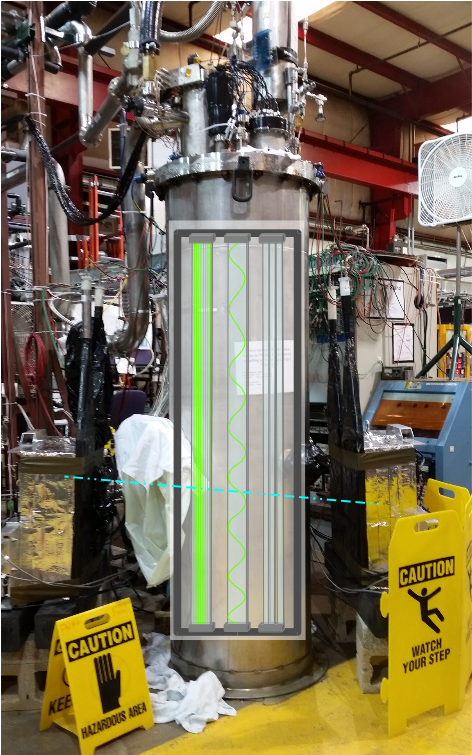}
    \vspace{-1.5em}
    \caption{TallBo, with hodoscopes located in the ``low'' position and an illustration of the light guides tested during Phase 1 superimposed.}
    \label{fig:TallBo4-Phase1-Layout}
    \vspace{-1em}
  \end{center}
\end{wrapfigure}

The TallBo dewar facility, Fig.~\ref{fig:TallBo4-Phase1-Layout}, is a large LAr cryostat at Fermilab's Proton Assembly Building capable of supporting several light guides in excess of 1.5 meters for side-by-side design performance testing. After mounting of light guides the dewar is evacuated to remove contamination from residual atmospheric gases. Once filled with ultra high purity LAr an LN2 condenser maintains a constant liquid level and internal pressure of 10~psig to prevent contamination of the LAr.

Scintillation signals were collected from SiPMs using SiPM Signal Processors (SSP), 150~MSPS waveform digitizers developed by the Argonne National Laboratory HEP Electronics Group for DUNE. Signal amplitude calibration data was collected in a self-triggered mode. Data used for performance analyses was collected from external triggers. Two modules from the CREST balloon-borne cosmic-ray detector~\cite{bib:CREST} flanked the TallBo dewar. Each module consists of an 8$\times$8 array of PMTs mounted on BaF scintillating crystals. To reject spurious photon activity in these modules paddles made from scintillating plastic were mounted in front of each array. This system provided a four-fold coincidence trigger to reject shower-like events and select single-track through-going muons for analysis.

\begin{figure}[ht]
  \begin{center}
    \begin{subfigure}{0.35\columnwidth}
      \centering
      \includegraphics[width=\columnwidth]{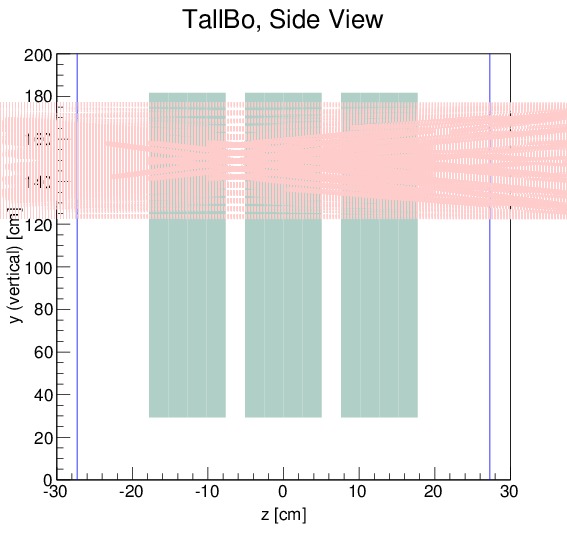}
      \vspace{-1.5em}\caption{}
      \label{fig:Phase2-TallBoYZ-High-Front}
    \end{subfigure}
    \hspace{0.05\columnwidth}
    \begin{subfigure}{0.35\columnwidth}
      \centering
      \includegraphics[width=\columnwidth]{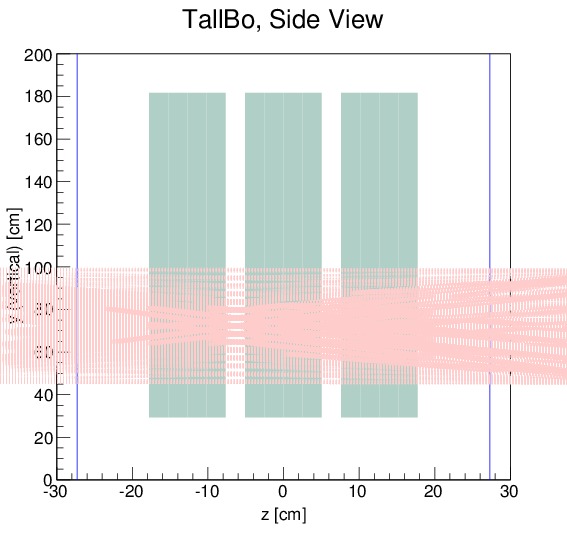}
      \vspace{-1.5em}\caption{}
      \label{fig:Phase2-TallBoYZ-Low-Front}
    \end{subfigure}
    \caption{(a) ``High'' and (b) ``low'' track selections.}
  \end{center}
  \vspace{-1em}
\end{figure}

Hodoscope triggers were accumulated in two positions, ``high'' and ``low.'' A single-track event is defined as one in which only one PMT detected a hit in each of the hodoscope arrays flanking the dewar during a four-fold coincidence. Furthermore, tracks with fewer than 15 photoelectrons detected in the 13~$\mu$s readout across all detectors were rejected. To facilitate comparison with light guides in the different fills, only tracks passing through the ``front'' region of the dewar were included. Complementary distributions from tracks through the ``back'' region were in agreement, indicating no substantial asymmetry between light guide faces. All tracks crossing from ``front'' to ``back'' or ``back'' to ``front'' were excluded to remove any tracks passing directly through a light guide. This track selection is illustrated in Figs.~\ref{fig:Phase2-TallBoYZ-High-Front} and \ref{fig:Phase2-TallBoYZ-Low-Front}.

\begin{figure}[ht]
  \begin{center}
    \begin{subfigure}{0.495\columnwidth}
      \centering
      \includegraphics[width=\columnwidth]{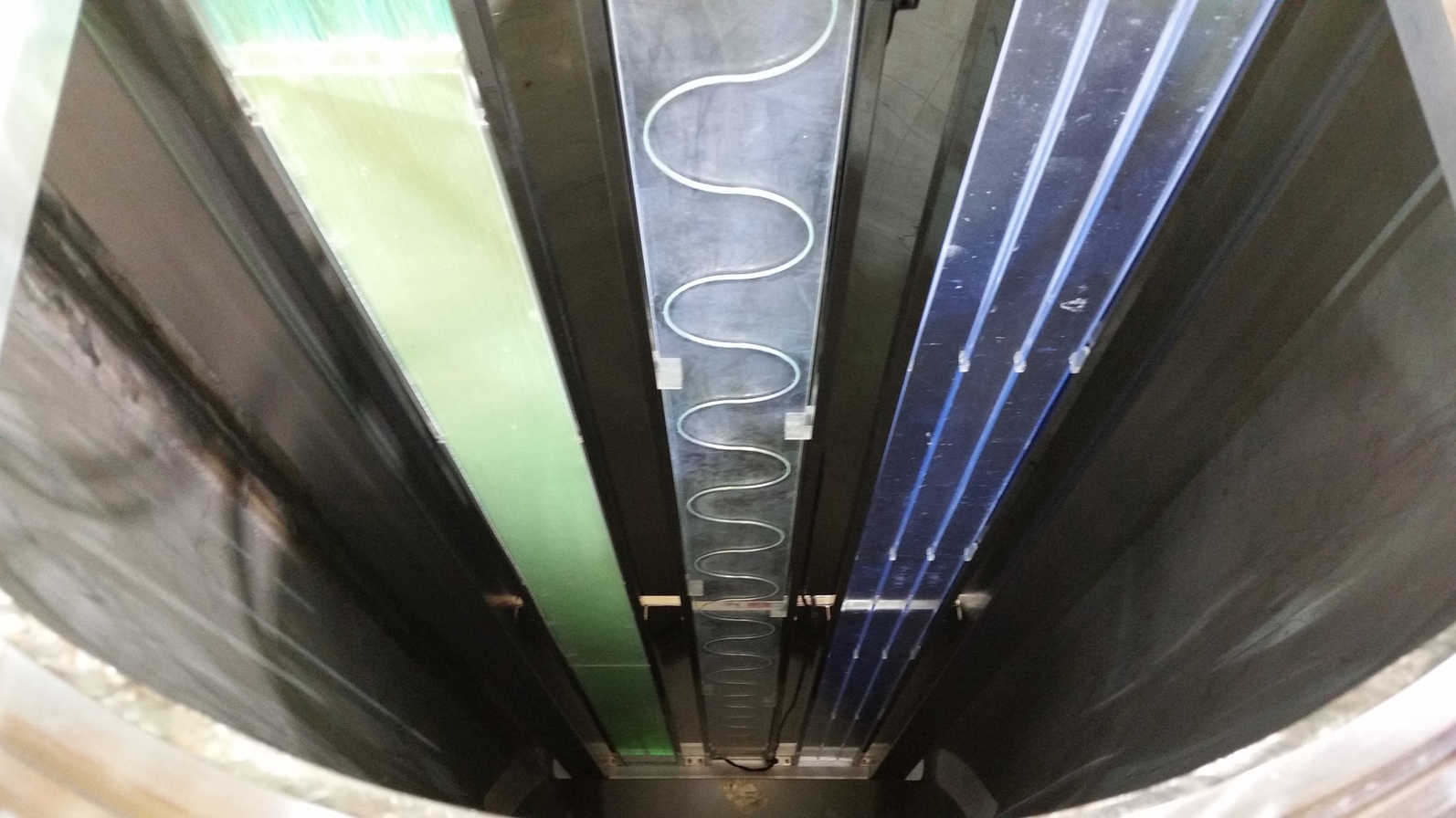}
      \vspace{-1.5em}\caption{}
      \label{fig:TallBo4-LightGuides-Phase1}
    \end{subfigure}
    \begin{subfigure}{0.495\columnwidth}
      \centering
      \includegraphics[width=\columnwidth]{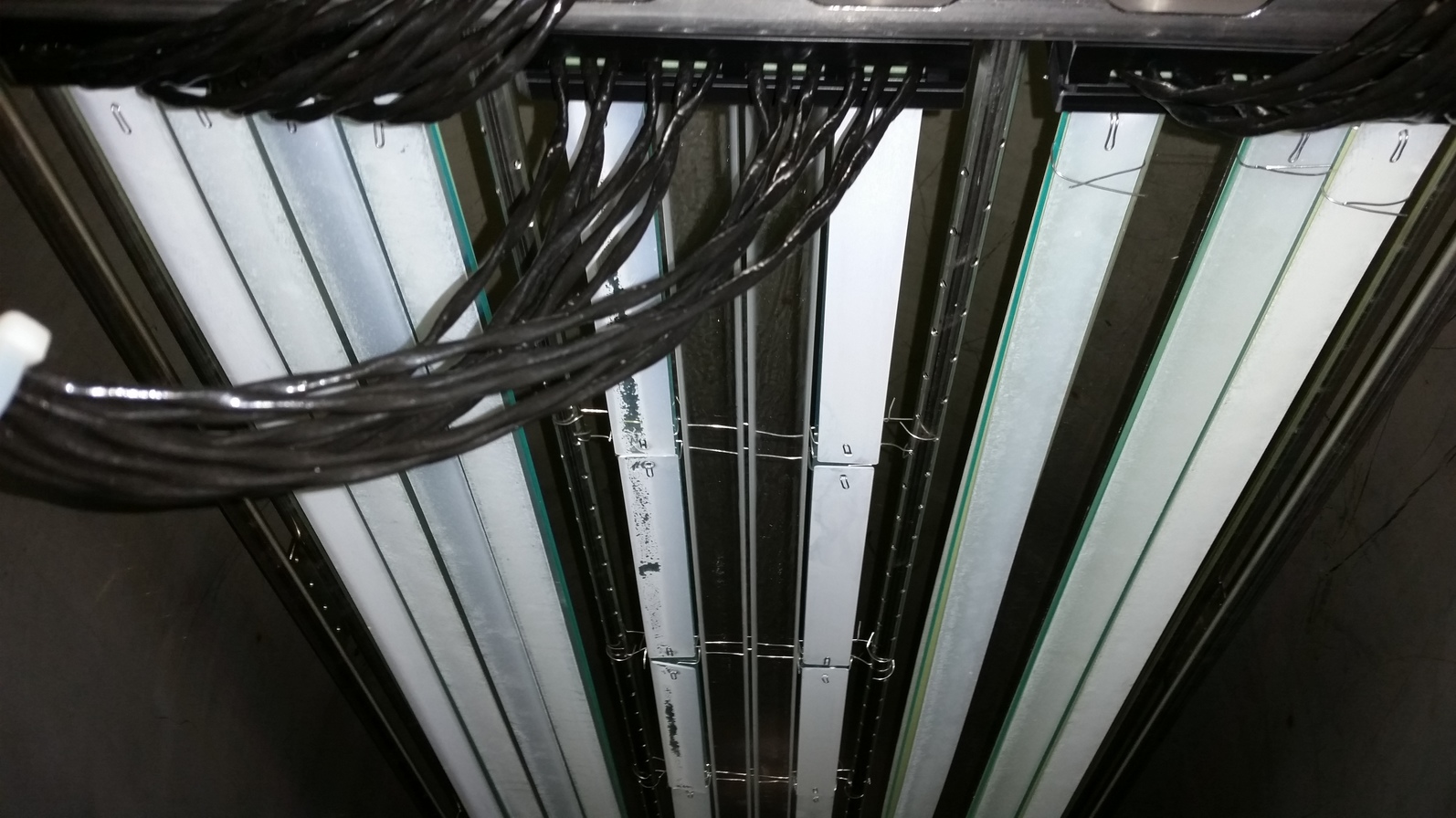}
      \vspace{-1.5em}\caption{}
      \label{fig:TallBo4-LightGuides-Phase2}
    \end{subfigure}
    \caption{The light guides compared during (a) Phase 1 and (b) Phase 2 of this experiment.}
  \end{center}
  \vspace{-1em}
\end{figure}

Three previous tests at TallBo have investigated several iterations of light guide designs, identifying the most promising candidate designs each time for focused continued refinement. The fourth such test was carried out between June 10 and August 19th, 2015. This experiment was the first deployment of $\sim$1.5~m light guide designs at the TallBo LAr facility and was aimed at careful measurement of both relative efficiencies and attenuation properties. To accommodate the various alternative designs the ``TallBo4'' experiment was divided into two LAr fills, designated ``Phase 1'' (Fig.~\ref{fig:TallBo4-LightGuides-Phase1}) and ``Phase 2'' (Fig.~\ref{fig:TallBo4-LightGuides-Phase2}).

Phase 1 of this experiment was carried out from June 10 through July 6. Three full-width (3~3/8'') light guides were tested side-by-side, comparing the baseline DUNE design to two alternatives intended to provide longer attenuation lengths. Phase 2 was carried out from July 28 through August 17. Twelve one-inch wide light guide designs were tested side-by-side. Nine of the light guides were of an alternate design consisting of a WLS-coated radiator plate in front of a wavelength shifter-doped light guide made from polystyrene (PS) or polyvinyl toluene (PVT). Also included were three baseline light guide designs.

Contamination by oxygen, water, and nitrogen were monitored during each LAr fill. After measurements were allowed to stabilize for several hours, oxygen was measured at less than 150 ppb, water at less than 15 ppb, and nitrogen at less than 90 ppb. There should be minimal impact on the scintillation signal at these contamination levels~\cite{bib:N2-Contam,bib:N2-Contam2,bib:O2-Contam,bib:H2O-Contam}.

The turbomolecular pump attached to the TallBo dewar malfunctioned during evacuation prior to the Phase 2 fill. Shrapnel and dust were ejected into the dewar. The photon detector modules were removed, the detectors and dewar were cleaned, and all equipment was visually inspected for damage. Although no indication of damage was observed and the experiment proceeded as planned, it is unknown what effects this event may have produced on the results.

\section{Light Guide Designs}

The version of the baseline module design manufactured by Indiana University (IU) (Fig.~\ref{fig:TallBo4-LightGuides-Phase1} {\em right}) consists of three acrylic light guides dip-coated with a solution of TPB dissolved in DCM at 0.6\% by weight. Scintillation light incident on each light guide is converted to blue visible photons. Some of these converted photons are captured by total internal reflection within the acrylic and transported to the readout end. Each of the three light guides is read out by an array of three SensL\footnote{sensl.com} C-series SiPMs at the top end. Two additional light guides of this manufacture were included in Phase~2 for comparison, along with one manufactured by MIT using a recently-improved technique expected to maintain a long attenuation length.

The Louisiana State University (LSU) design (Fig.~\ref{fig:TallBo4-LightGuides-Phase1} {\em center}) consists of a bundle of three wavelength shifting fibers embedded in an acrylic panel. The panel is then dip-coated in a manner similar to the baseline design with a solution TPB and acrylic dissolved in toluene. Scintillation light incident on the panel is converted to blue visible photons. Blue light trapped inside the panel is captured directly or after reflections by the fiber bundle where it is converted to green light and transported to the ends of the fibers. Each end of the fiber bundle is guided onto one SensL B-series SiPMs to accomplish double-ended readout.

The Colorado State University (CSU) design (Fig.~\ref{fig:TallBo4-LightGuides-Phase1} {\em left}) consists of an array of wavelength shifting fibers behind a polystyrene radiator plate hand-painted on the outer face with a solution of TPB dissolved in toluene. Scintillation light incident on the radiator is converted to blue visible photons. Roughly half of this blue light impinges on fibers where it is converted to green light and transported to the readout end. The fibers are guided onto six SensL B-series SiPMs at the top end.

\begin{figure}[ht]
  \begin{center}
    \includegraphics[width=0.7\columnwidth]{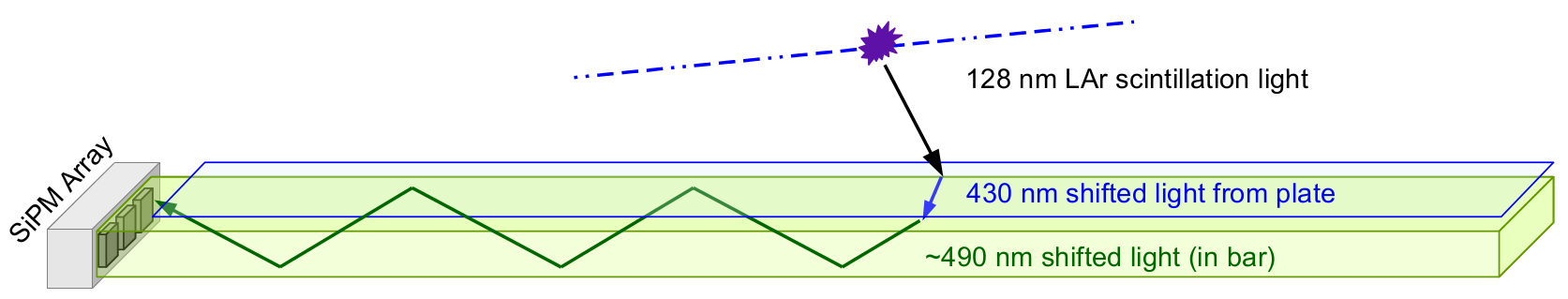}
    \caption{Light guide design consisting of a TPB-coated radiator plate in front of a polystyrene or polyvinyl toluene light guide doped with the EJ-280 WLS.}
    \label{fig:PlateAndLightGuide}
  \end{center}
  \vspace{-1em}
\end{figure}

An alternate light guide designed and manufactured by IU, Fig.~\ref{fig:PlateAndLightGuide}, consists of a WLS-coated radiator plate in front of a light guide manufactured by Eljen\footnote{www.eljentechnology.com/index.php/component/content/article/30-the-community/91-ej-280} using polystyrene (PS) or polyvinyl toluene (PVT) doped with a green-to-blue wavelength shifter (EJ-280). Scintillation light incident on the plate is converted to blue light. Approximately half of this blue light is incident on the EJ-280 light guide where it is converted to green with about 85\% efficiency. A portion of this converted light is transported by total internal reflection to the top where three SensL C-series SiPMs read out each one-inch-wide light guide. A number of substrates, coating solutions, and coating methods are being explored. Several variations of this design were deployed during Phase 2 of this experiment (Fig.~\ref{fig:TallBo4-LightGuides-Phase2}).

\section{Performance Comparisons}

\subsection{Phase 1 Design Comparisons}

\begin{figure}[ht]
  \begin{center}
    \begin{subfigure}{0.325\columnwidth}
      \centering
      \includegraphics[width=\columnwidth]{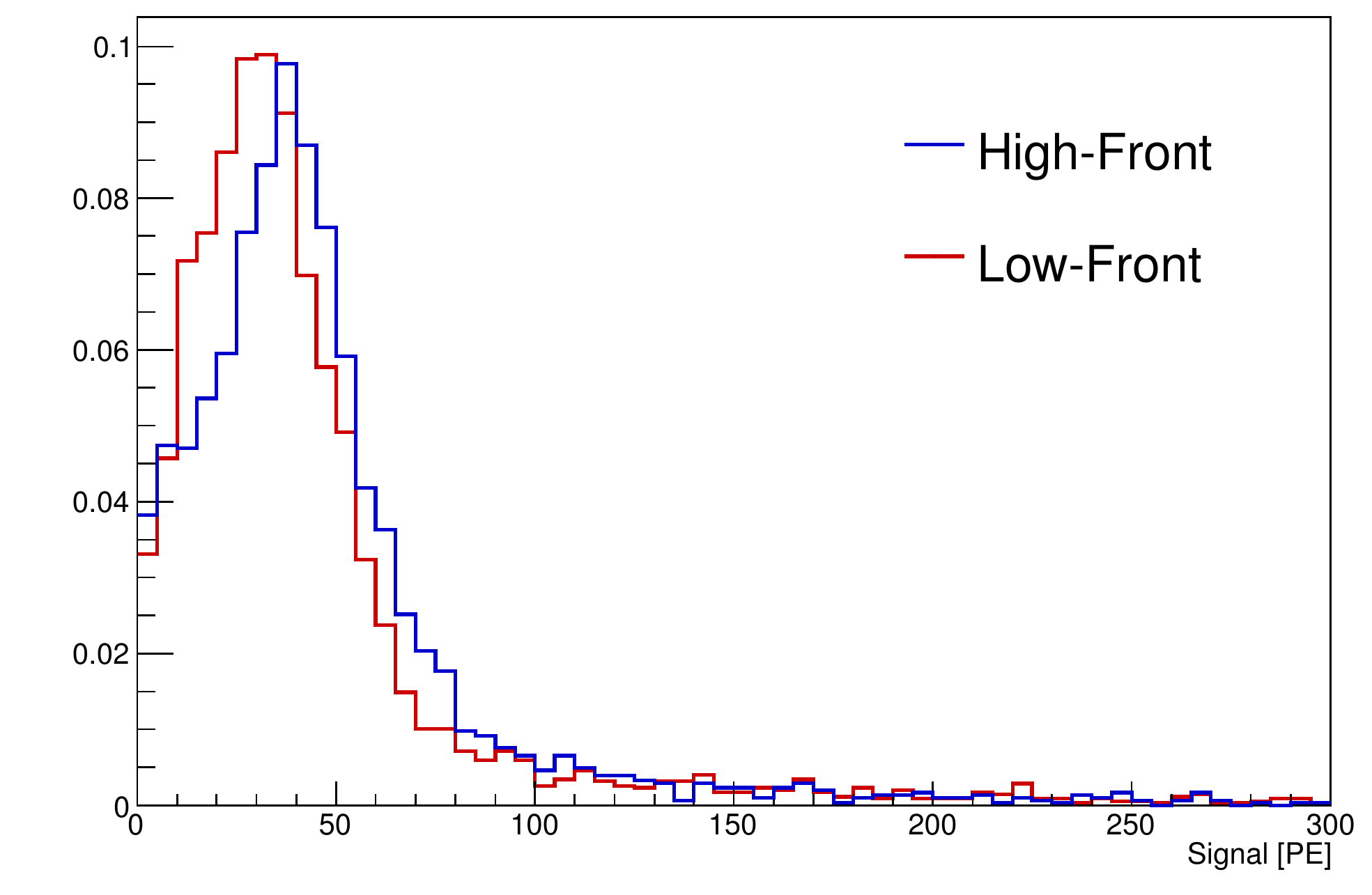}
      \vspace{-1.5em}\caption{}
      \label{fig:TrackSigs-HL-CSU}
    \end{subfigure}
    \begin{subfigure}{0.325\columnwidth}
      \centering
      \includegraphics[width=\columnwidth]{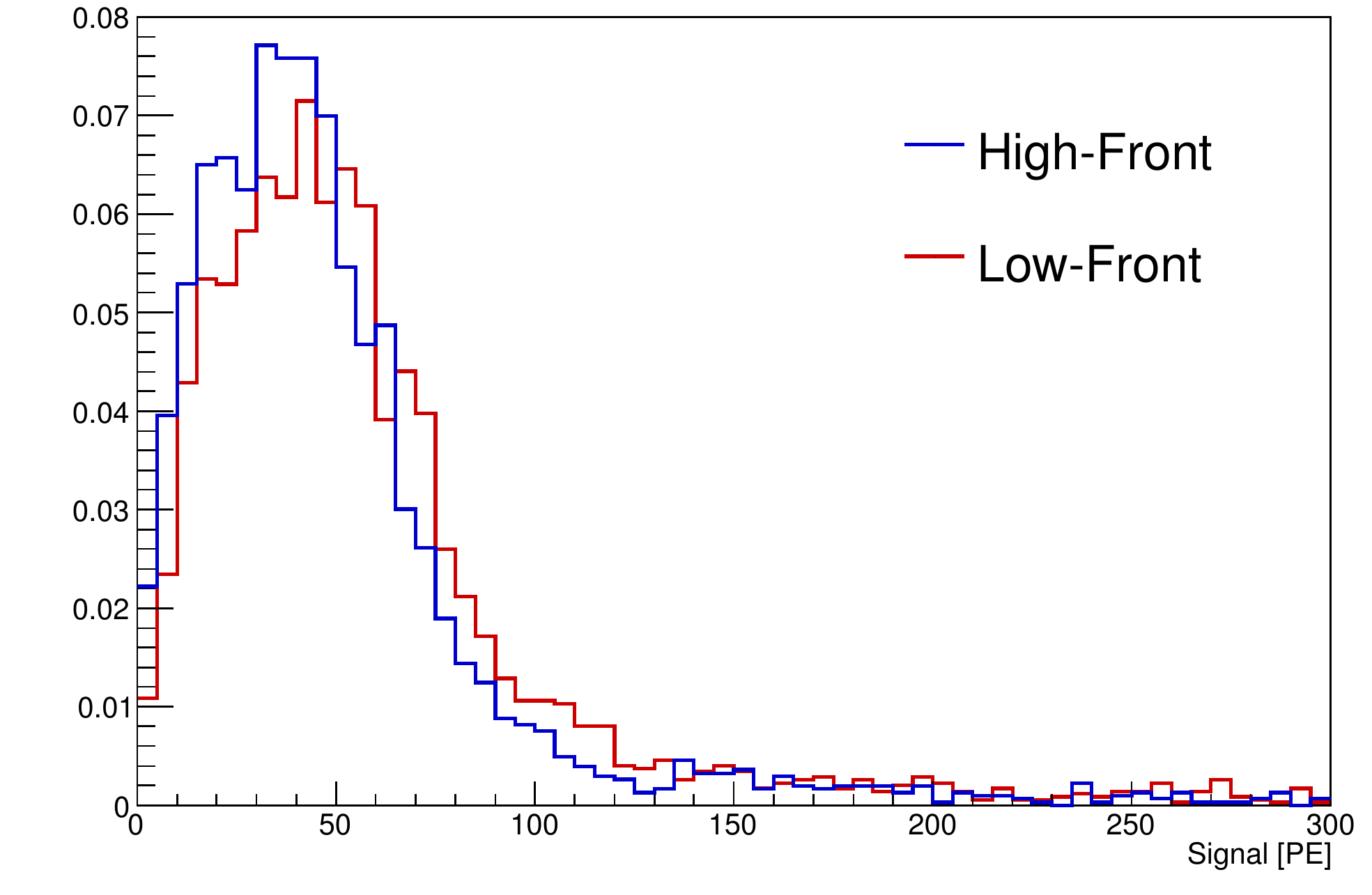}
      \vspace{-1.5em}\caption{}
      \label{fig:TrackSigs-HL-LSU}
    \end{subfigure}
    \begin{subfigure}{0.325\columnwidth}
      \centering
      \includegraphics[width=\columnwidth]{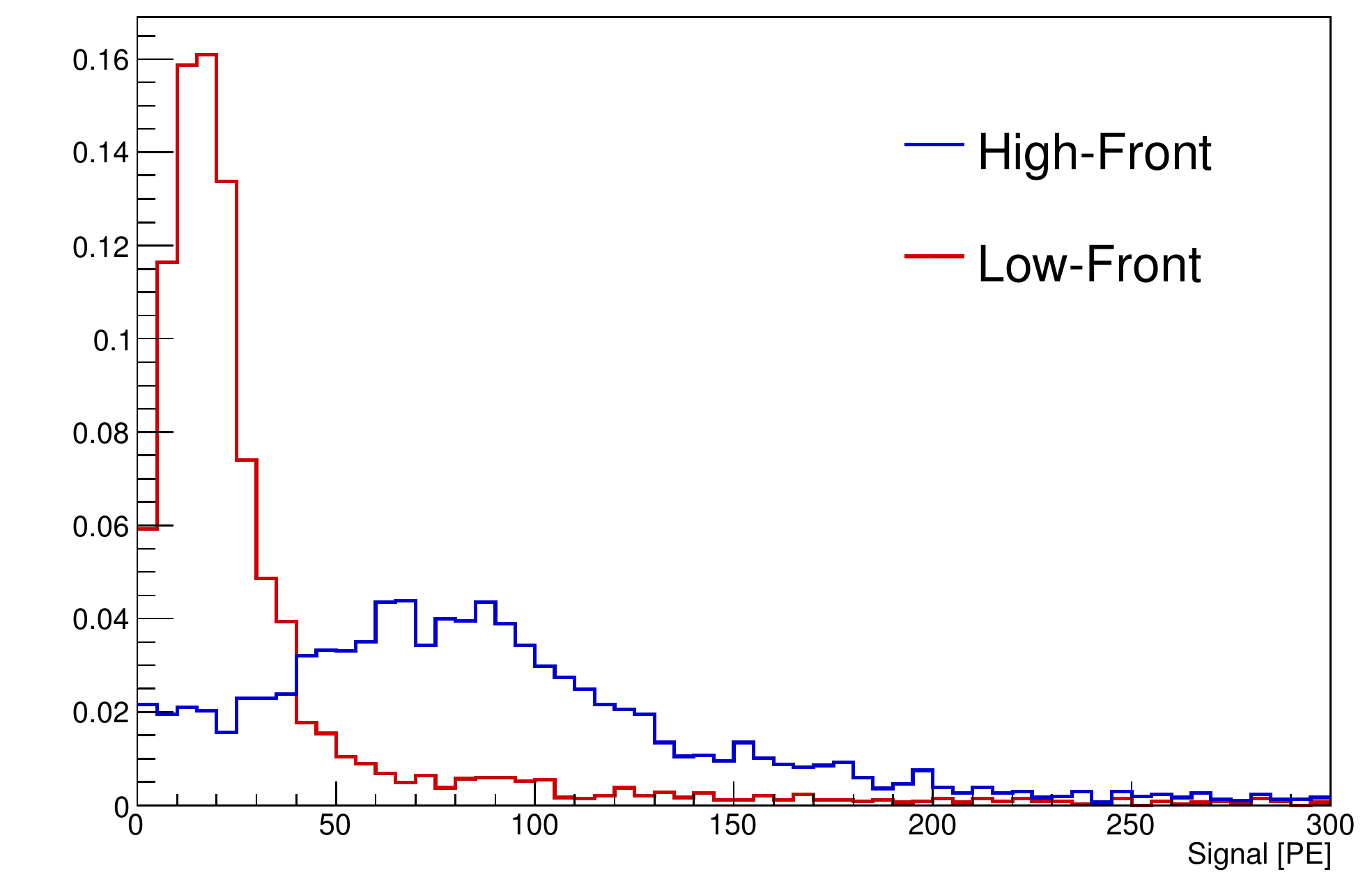}
      \vspace{-1.5em}\caption{}
      \label{fig:TrackSigs-HL-IU}
    \end{subfigure}
    \caption{Normalized distribution of signal amplitudes measured by (a) the CSU fiber paddle, (b) the LSU fiber paddle, and (c) the IU dip-coated paddle.}
  \end{center}
  \vspace{-1em}
\end{figure}

Figures~\ref{fig:TrackSigs-HL-CSU}--\ref{fig:TrackSigs-HL-IU} show distributions of the total signal detected in the 13~$\mu$s readout from single-track events, calibrated to photoelectrons (pe) and summed over all SiPMs reading out each module. The distributions are normalized to unit area. The blue and red histograms compare signals from the ``high'' and ``low'' selections, respectively.

Direct comparison of these distributions gives an approximate measure of the attenuation properties of each design. The CSU and LSU fiber-based designs showed minimal attenuation loss between the two track selections. The IU dip-coated design collected the most light from tracks near its readout end but showed the most substantial attenuation loss.

\subsection{Phase 2 Design Comparisons}

\begin{figure}[ht]
  \begin{center}
    \begin{subfigure}{0.325\columnwidth}
      \centering
      \includegraphics[width=\columnwidth]{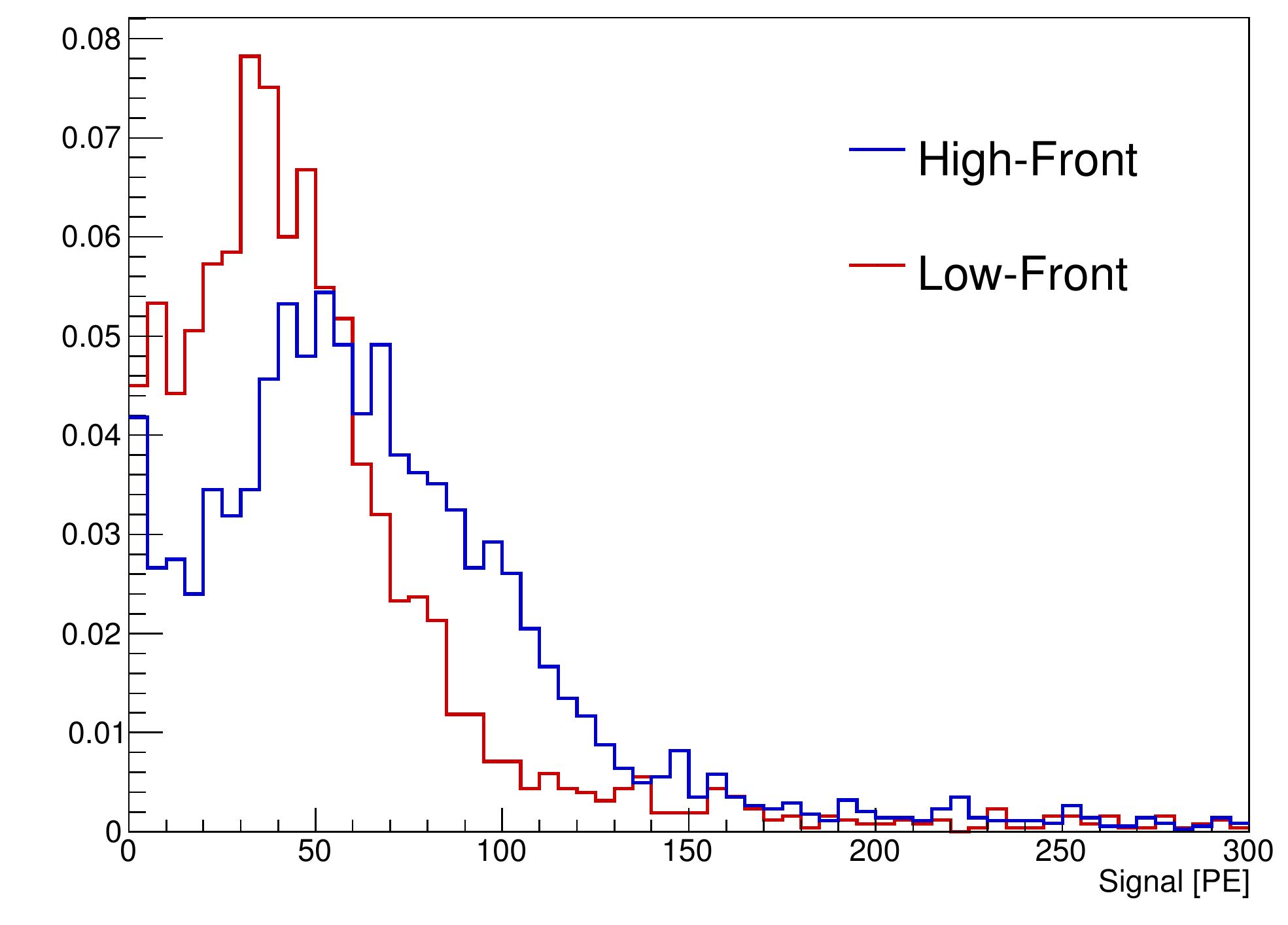}
      \vspace{-1.5em}\caption{}
      \label{fig:TrackSigs-HL-C-0}
    \end{subfigure}
    \hspace{0.03\columnwidth}
    \begin{subfigure}{0.325\columnwidth}
      \centering
      \includegraphics[width=\columnwidth]{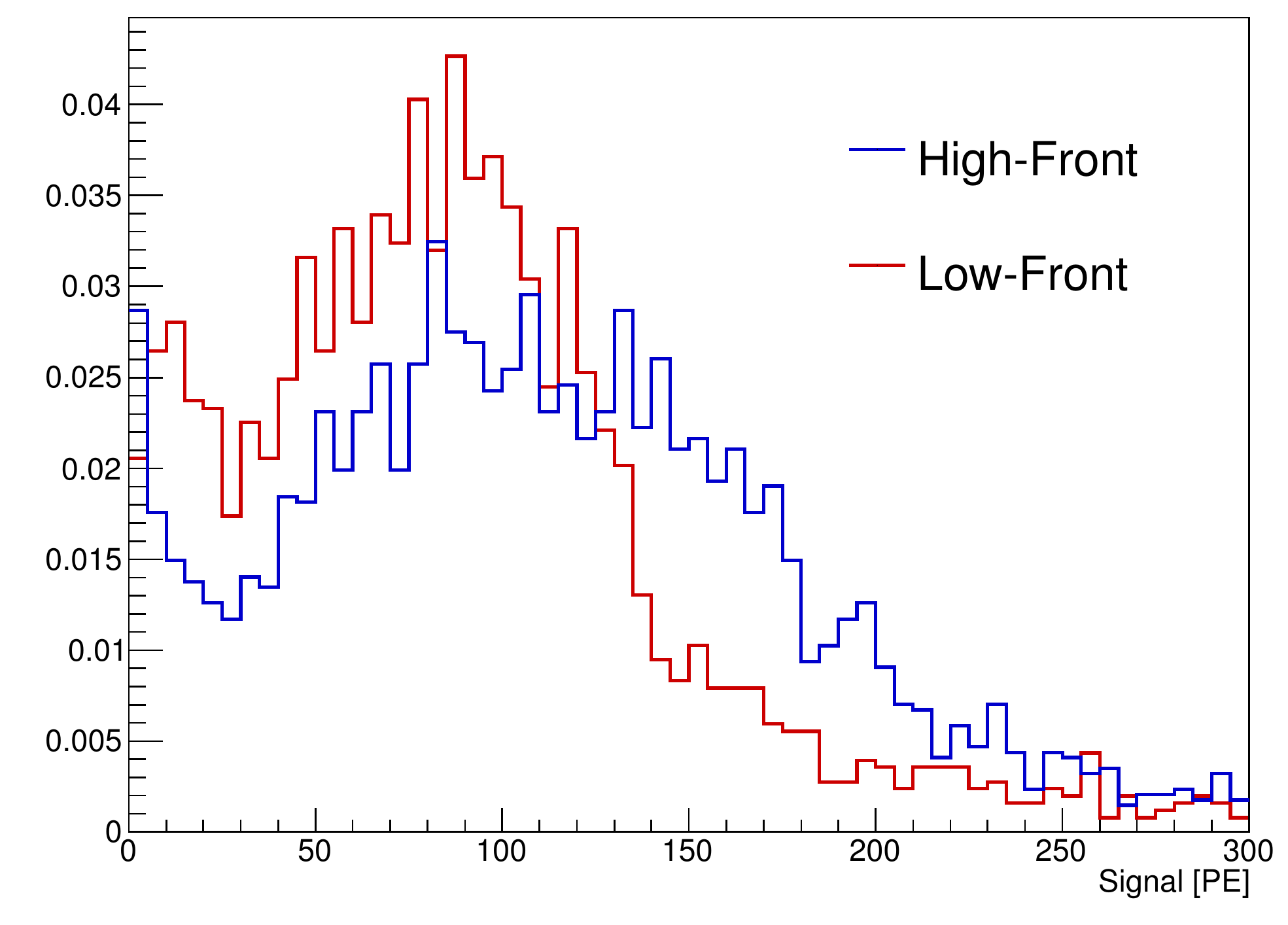}
      \vspace{-1.5em}\caption{}
      \label{fig:TrackSigs-HL-C-1}
    \end{subfigure}
    \\\vspace{1em}
    \begin{subfigure}{0.325\columnwidth}
      \centering
      \includegraphics[width=\columnwidth]{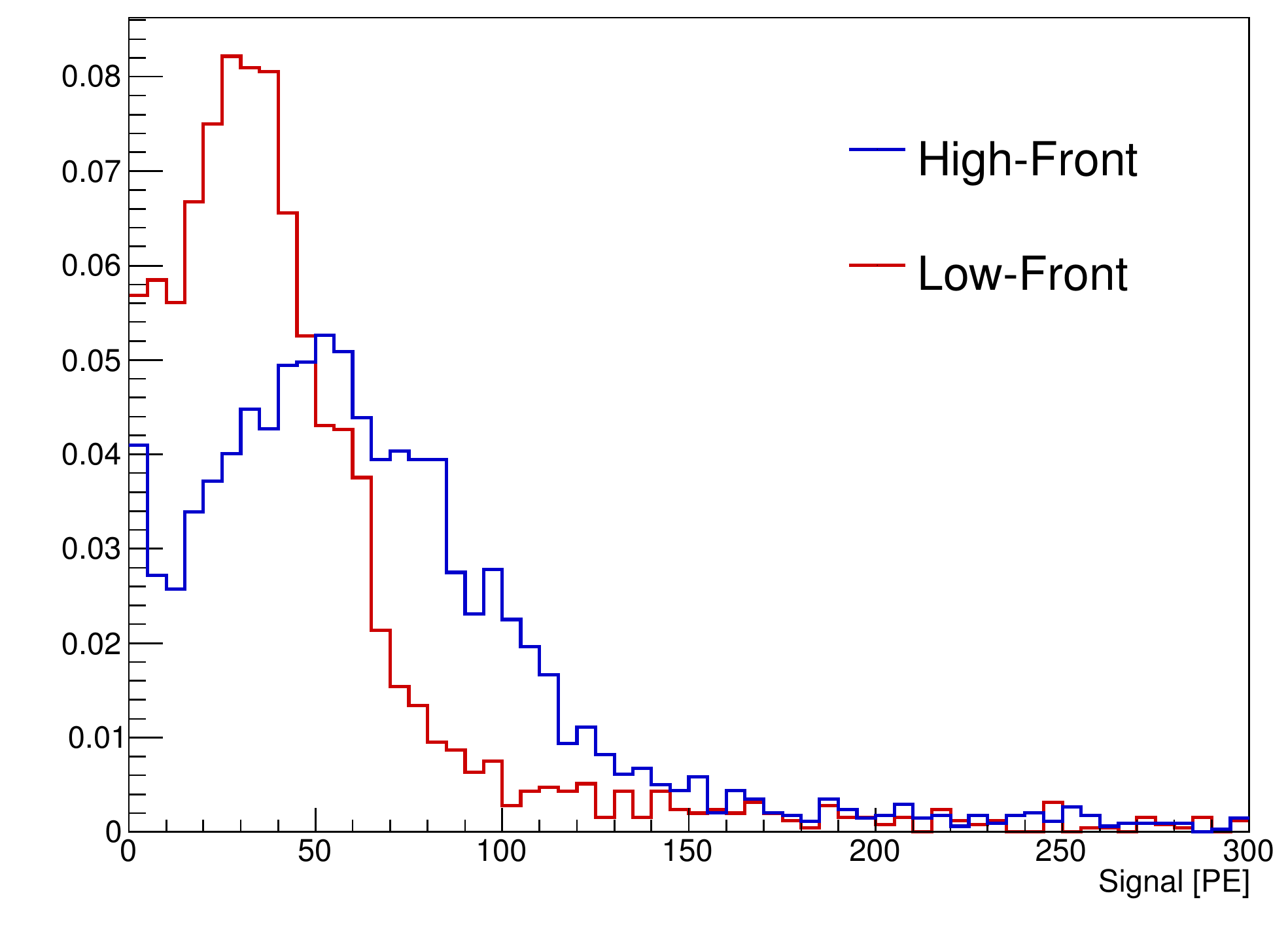}
      \vspace{-1.5em}\caption{}
      \label{fig:TrackSigs-HL-C-2}
    \end{subfigure}
    \hspace{0.03\columnwidth}
    \begin{subfigure}{0.325\columnwidth}
      \centering
      \includegraphics[width=\columnwidth]{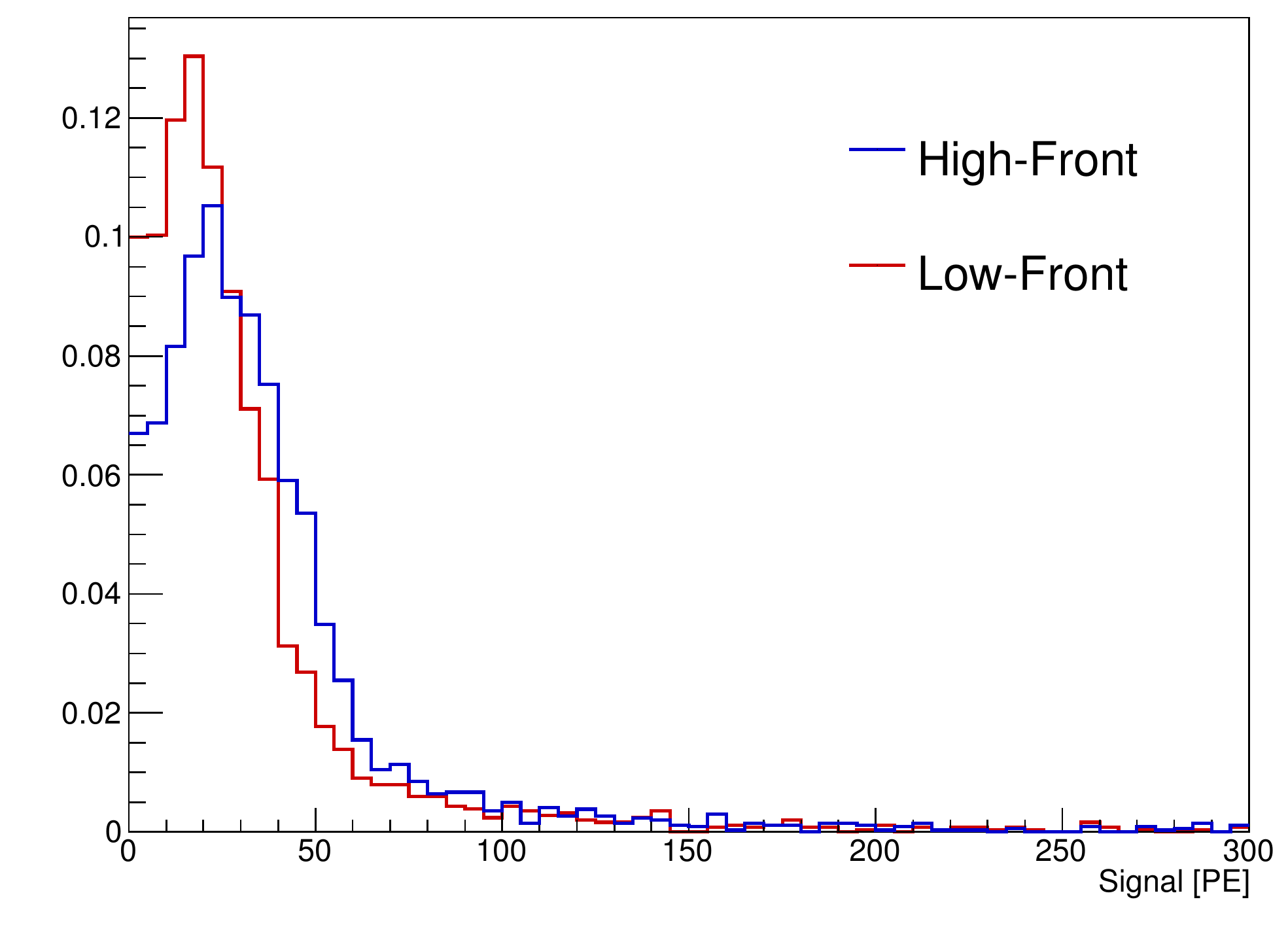}
      \vspace{-1.5em}\caption{}
      \label{fig:TrackSigs-HL-C-3}
    \end{subfigure}
    \caption{Normalized distribution of signal amplitudes measured by (a) WLS PVT with TPB-coated acrylic plate, (b) Dip-coated acrylic light guide manufactured by MIT, (c) WLS PS with TPB-coated acrylic plate, (d) WLS PS with bis-MSB-coated acrylic plate.}
  \end{center}
  \vspace{-1em}
\end{figure}

Figs.~\ref{fig:TrackSigs-HL-C-0}--\ref{fig:TrackSigs-HL-C-3} compare three of the EJ-280 light guides with WLS-coated plates and one dip-coated acrylic light guide manufactured by MIT. The alternative designs using TPB-coated plates with EJ-280 light guides represent an improvement over the baseline design from Phase 1 by a factor of two.\footnote{
Data collected from the baseline design during Phase 1 represents the total light detected simultaneously by three one-inch light guides and is roughly three times the signal expected for a single one-inch light guide. Extrapolating from the Phase 1 ``high'' data to the Phase 2 ``high'' data, the mean signal detected in this track selection would be $\sim$25~pe.}
Light guides using plates coated with TPB outperformed those using bis-MSB, indicating that the latter wavelength shifter may be more difficult to bind to a substrate using the methods tested here. No discernible difference was found between the PS- and PVT-based EJ-280 light guides. Acrylic and fused silica were tested as plate substrates and both performed equally well.

The new recipe from MIT for manufacturing light guides using the baseline dip-coat method has produced a substantial improvement over previous manufacturing methods and represent an improvement over the alternative designs in Phase 2 by almost a factor of 2. Attempts are underway at IU to produce light guides using this new manufacturing method in order to investigate the reproducibiliy of light guides with this level of performance.

The effect of attenuation in these light guides can be estimated from the reduction of the mean signal amplitude between the two track selections. The central position of the track selections differ by about 65 cm. The means were estimated by Gaussian fits to the core of each histogram. The results are collected in Table~\ref{tab:AttenLengths}. It is likely that these figures underestimate the attenuation length of each light guide. The distributions represent signals from tracks spanning a wide region along and perpendicular to the light guides inside a cylindrical dewar. Improved measurements will come from scans of the light guide with a movable Am-241 source in the IU dewar facility and by comparison between the track data and Monte Carlo simulations. Nonetheless, the results presented here indicate substantial improvements in attenuation over the baseline design.

\begin{table*}[ht]
  \begin{center}
    \caption{Approximate attenuation lengths}\vspace{-0.5em}
    \label{tab:AttenLengths}
    \begin{tabular}{l c}
      \hline
      \hline
      Light guide & Attenuation Length \\
      \hline
      WLS PVT, TPB Acrylic Plates & $>$155 cm \\
      Dip-Coated Acrylic (New MIT Recipe) & $>$185 cm \\
      WLS Polystyrene, TPB Acrylic Plates & $>$110 cm \\
      WLS Polystyrene, bis-MSB Acrylic Plates & $>$155 cm \\
      \hline
      \hline
    \end{tabular}
  \end{center}
  \vspace{-1em}
\end{table*}

\section{Xenon-Doped Liquid Argon}

After collection of the Phase 2 comparison data was completed a series of injections of xenon into the LAr was carried out. Collisional interactions between $\text{Ar}_2^*$ eximers and dissolved atomic xenon would effectively result in triggered emission of energy from triplet $\text{Ar}_2^*$ eximers in the form of 175~nm photons on a timescale associated with random molecular motion within the liquid~\cite{bib:XeAr}. This process may produce a faster total signal and present the possibility for more efficient detection of the longer wavelength light.

\begin{figure}[ht]
  \begin{center}
    \begin{subfigure}{0.4\columnwidth}
      \centering
      \hspace{-0.095\columnwidth}      \includegraphics[width=\columnwidth]{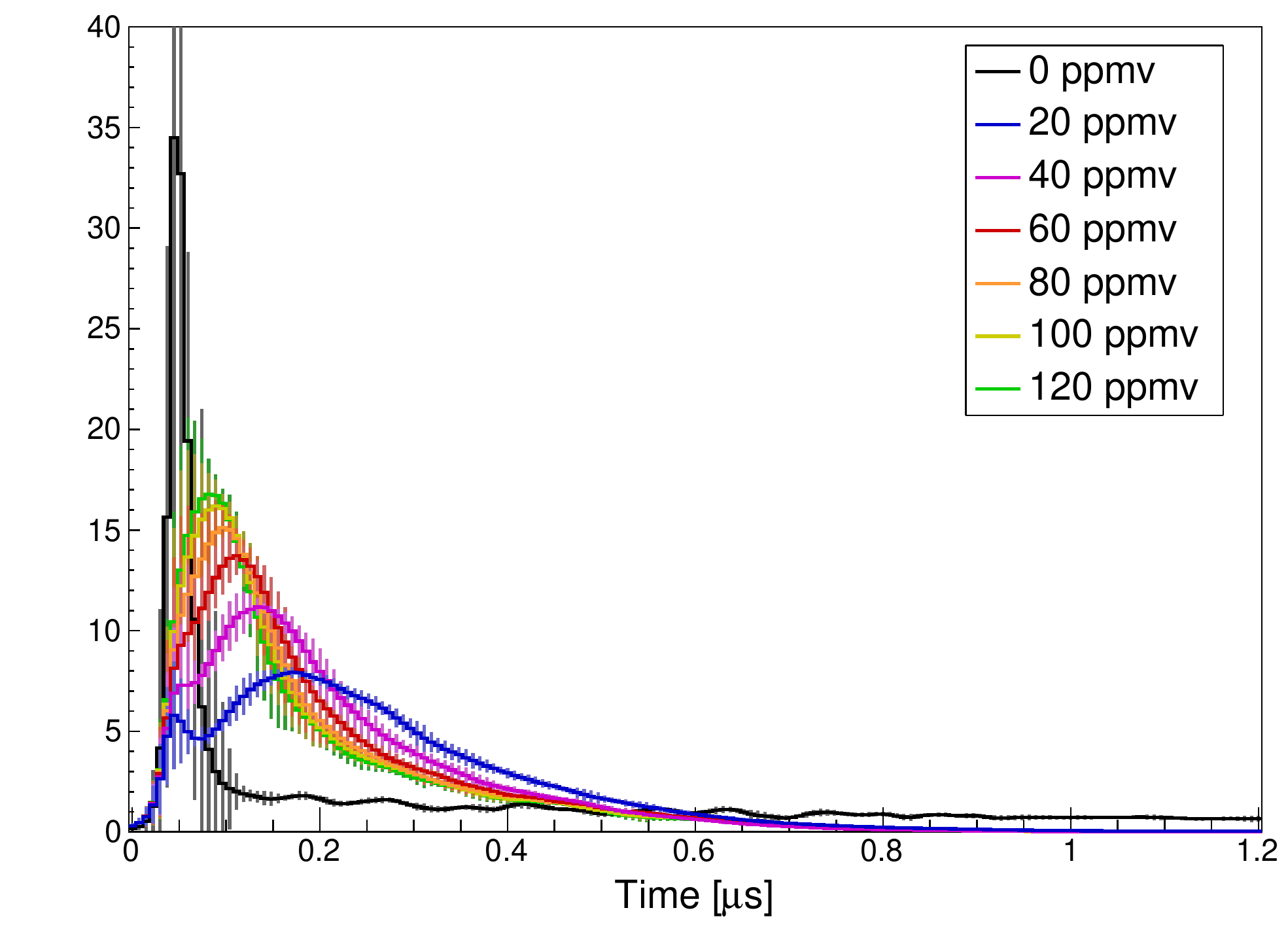}
      \caption{}
      \label{fig:XenonSignals}
    \end{subfigure}
    \hspace{0.03\columnwidth}
    \begin{subfigure}{0.4\columnwidth}
      \centering
      \hspace{-0.095\columnwidth}      \includegraphics[width=\columnwidth]{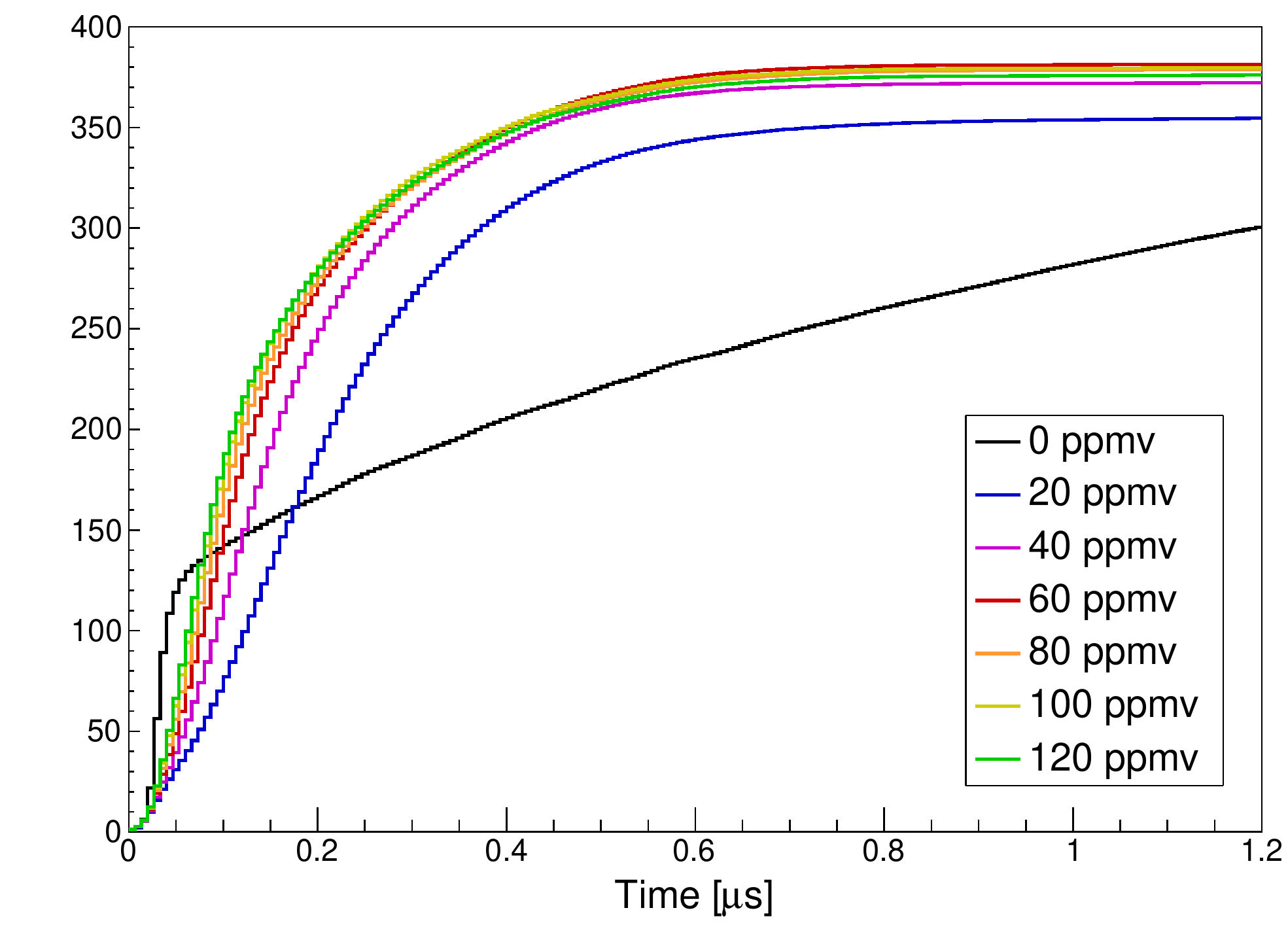}
      \caption{}
      \label{fig:XenonSignals-Int}
    \end{subfigure}
    \caption{(a) Time-dependent structure of the xenon-doped LAr signal from cosmic-ray muons in the first 1.2~$\mu$s at various concentrations. (b) Cumulative scintillation signal from xenon-doped LAr.}
  \end{center}
  \vspace{-1em}
\end{figure}

Gaseous xenon was mixed with gaseous argon, heated, and injected into the LAr in increments of approximately 20~ppm (by volume). Data was collected in a self-triggered mode whereby 11~$\mu$s waveforms were recorded from scintillation activity registering a prompt amplitude greater than $\sim$7~pe. The average waveforms collected by each SiPM were then deconvolved using the average SiPM single-microcell response to recover the time-dependent structure of the detected scintillation signal. This analysis is similar to that used to study the scintillation signal from cosmic-ray muons in pure LAr during a previous experiment~\cite{bib:LArScint-CRMuons}. The data presented in Figs.~\ref{fig:XenonSignals} and \ref{fig:XenonSignals-Int} were recorded using one SiPM mounted on a light guide consisting of a PVT EJ-280 light guide with TPB-coated acrylic radiator plates.

A striking change in the time structure of the scintillation signal occurs at concentrations as low as 20 ppmv (blue) compared to the signal from pure LAr (black). The long tail of triplet state emission is replaced with a broad signal peaking at $\sim$150~ns. Similar effects have been reported for scintillation in xenon-doped LAr induced by electrons and neutrons~\cite{bib:XeAr}. By 200~ns more light has been collected from the 20~ppmv xenon-doped LAr signal than from the undoped signal, and this effect becomes more pronounced as the xenon concentration increases. The prompt signal appears diminished in Fig.~\ref{fig:XenonSignals}, although the self-triggered threshold may contribute to this effect if the efficiency to detect scintillation light is different at 128~nm and 175~nm. Further studies using externally-triggered cosmic-ray data will help resolve these ambiguities.

\clearpage
\section{Conclusions}

Multiple VUV-sensitive light guide designs have been proposed for the DUNE photon detector system. Direct performance comparisons of $\sim$1.5 m light guides have been successfully carried out in a controlled ultra-high purity LAr environment using signals from cosmic-ray muons.

The version of the baseline design manufactured by MIT and the EJ-280 light guides with TPB-coated radiator plates proved to be the most promising designs in this experiment. Further tests are planned to explore the repeatability of the MIT recipe and manufacturing technique for the baseline design. The EJ-280 light guides maintain a long attenuation length when immersed in LAr. A variety of radiator plate substrates and coating methods are being explored at IU to improve the detection efficiency of this alternative design. The results from the baseline light guides manufactured by IU and from several previous design tests indicate the need to explore decoupling of the VUV conversion from transport in the light guide. All three such alternative designs tested in this experiment demonstrated minimal attenuation loss.

The analysis of scintillation light from cosmic-ray muons in xenon-doped LAr indicates it is possible to shorten the triplet $\text{Ar}_2^*$ signal to less than 200~ns. If this altered emission signal occurs at a longer wavelength, detection efficiency may also be improved. These two factors may be of value to a large LAr TPC like the DUNE far detector and warrant further study.

A return to the TallBo facility is anticipated in early 2016. Refined versions of the baseline design and the alternatives using WLS-coated plates with EJ-280 light guides will be deployed and their response to cosmic-ray muons will be studied. The xenon injection will also be repeated in order to collect high statistics from external triggers, investigate effects at lower xenon concentrations, and distinguish differences in the time-resolved scintillation signal at 128~nm from scintillation at longer wavelengths.

\section*{Acknowledgments}

This work was partially supported by the Office of High Energy Physics of the DOE with grant DE-SC0010120 to Indiana University and by Brookhaven National Laboratory with grant 240296-A3 to Indiana University.  Operated by Fermi Research Alliance, LLC under Contract No. DE-AC02-07CH11359 with the United States Department of Energy. The author wishes to thank the many people who helped make this work possible. At IU: B.~Adams, B.~Baptista, B.~Baugh,  M.~Gebhard, B.~Howard, M.~Lang, S.~Mufson, J.~Musser, P.~Smith, J.~Urheim.  At MIT:  L.~Bugel, J.~Conrad, B.~Jones, Z.~Moss, M.~Toups, T.~Wongjirad.  At Fermilab: R.~Davis, M.~Geynisman, K.~Hardin, M.~Johnson, W.~Miner, S.~Pordes, B.~Rebel, M.~Ruschman. At ANL: J.~Anderson, P.~DeLurgio, G.~Drake, V.~Guarino, A.~Kreps, M.~Oberling.  At Louisiana State: T.~Kutter, J.~Insler. At Colorado State: D.~Adams, N.~Buchanan, F.~Craft, T.~Cummings, J.~Jablonski, D.~Warner, R.~Wasserman.

\end{document}